\documentclass[11pt,a4paper,twoside]{article}
\usepackage[utf8]{inputenc}
\usepackage[english]{babel}
\usepackage{amsmath,amsfonts,amssymb}
\usepackage{graphicx}
\usepackage{amsthm} 
\usepackage{empheq}
\usepackage{tensor} 
\usepackage{hyperref} 
\usepackage{subfig} 
\usepackage{xcolor}
\usepackage{cleveref}
\usepackage{bm} 
\usepackage[most]{tcolorbox}
\usepackage{afterpage} 
\usepackage{todonotes} 
\usepackage{lscape} 
\usepackage{booktabs} 
\usepackage{esint}  
\usepackage{footmisc} 
\usepackage{transparent}

\definecolor{graylight}{cmyk}{.30,0,0,.67} 

\numberwithin{equation}{section}

\newcommand{\T}[2]{\tensor{#1}{#2}}
\newcommand{\p}{\partial}
\newcommand{\g}{\gamma}

\newcommand{\C}[1]{\mathcal{C}l_{#1}}
\newcommand{\hl}[1]{\colorbox{yellow}{#1}}
\newcommand{\hf}{\frac{1}{2}} 
\newcommand{\dd}{\mathrm{d}}

\crefname{equation}{equation}{equations}
\Crefname{equation}{Equation}{Equations}
\crefrangelabelformat{equation}{(#3#1#4--#5#2#6)}

\crefmultiformat{equation}{equations (#2#1#3}{, #2#1#3)}{#2#1#3}{#2#1#3}
\Crefmultiformat{equation}{Equations (#2#1#3}{, #2#1#3)}{#2#1#3}{#2#1#3}

\title{Differential Forms vs Geometric Algebra: The quest for the best geometric language}
\date{\today \\ University of Heidelberg}
\author{Pablo Bañón Pérez and Maarten DeKieviet}

\begin{document}
	
\maketitle

\begin{abstract}
	Differential forms is a highly geometric formalism for physics used from field theories to General Relativity (GR) which has been a great upgrade over vector calculus with the advantages of being coordinate-free and carrying a high degree of geometrical content.

	In recent years, Geometric Algebra appeared claiming to be a unifying language for physics and mathematics with a high level of geometrical content. Its strength is based on the unification of the inner and outer product into a single geometric operation, and its easy interpretation. 
	
	Given their similarities, in this article we compare both formalisms side-by-side to narrow the gap between them in literature. We present a direct translation including differential identities, integration theorems and various algebraic identities.
	
	As an illustrative example, we present the case of classical electrodynamics in both formalism and finish with their description of GR.
\end{abstract}

\tableofcontents

\section{Introduction}

The evolution of different refinements and generalizations of calculus has provided physicists with improved tools to deal with physical problems. The most common formalism used in physics is vector calculus, developed by J. W. Gibbs and O. Heaviside near the end of the 19th century, with most of today's notation and terminology established in the 1901 book, \textit{Vector Analysis}. However, the widespread adoption of this formalism was not based on a critical assessment of its computational power, but on a series of particular historical events \cite{Vector_War_Chappell_2016}.

Many of the limitations of vector calculus were addressed by the differential forms formalism, created by Cartan in 1899\cite{Cartan_1899}. Differential forms are particularly suited for analyzing fields on manifolds and vector spaces, providing a more flexible and powerful framework than traditional vector calculus or tensor calculus. Differential forms offer coordinate independence, simplify integration over manifolds, and provide a unified approach to Maxwell's equations and gauge theory. This formalism naturally incorporates the concept of gauge invariance, as differential forms are inherently coordinate-independent and the exterior derivative respects gauge transformations. This alignment ensures that physical laws expressed in terms of differential forms remain invariant under gauge transformations, a fundamental principle in many areas of physics such as electromagnetism, the weak and strong nuclear forces, and general relativity. A very didactic, albeit a bit outdated, comparison of vector calculus and differential forms was presented in \cite{Schleifer_1983}.

In recent years, Geometric Algebra (GA)\footnote{Also called Clifford Algebra.} has emerged as a new alternative formalism. Originally developed by William Kingdon Clifford in 1870, it experienced a renaissance through the work of David Hestenes and the Cambridge group led by Anthony and Joan Lasenby, Chris Doran, and others \cite{Hestenes_NFCM,Hestenes_GC,Doran2013}.

The Geometric Algebra (GA) and Geometric Calculus (GC) formalisms provide a unified mathematical language that combines scalars, vectors, and higher-dimensional objects called multivectors into a single framework \cite{Hestenes_A_Unified_Language}, simplifying calculations by eliminating the need to switch between different mathematical objects. GA encompasses complex numbers and analytic theory, quaternions, exterior algebra, and even the Pauli and Dirac algebras, integrating them into a coherent system. This unification allows for a more intuitive and efficient handling of geometric transformations, rotations, and reflections. GC extends GA to include differentiation and integration, respecting the geometric structure of space-time and making it particularly well-suited for formulating physical laws in a coordinate-free manner. Additionally, GA and GC incorporate tensor calculus and differential forms, offering powerful tools for dealing with fields and differential equations.

Because both differential forms and GA are based on exterior calculus, they have many similarities, but also differences. In this paper, we will compare them side-by-side to create a bridge between both camps in the literature. Our goal is to introduce GA and its power to those familiar with differential forms, and equip those with knowledge of GA with the translation tools necessary to read and communicate in the language of differential forms.

Because it is the lesser-known language, we will start with a presentation of GA and its evolution into GC, followed by a section where we compare it with differential forms, including the most common algebraic identities, and differential and integral theorems. Then, we will present the description of relativistic electrodynamics as an example, before providing an overview of their similarities and differences in the treatment of curved manifolds and General Relativity (GR).

We find that the main issues with differential forms are the absence of mixed-grade elements and a complicated intuitive geometrical interpretation. GC addresses both problems, unifies various differential operators, and expands the applicability of the generalized Stokes' theorem.

\section{Theory of geometric calculus} 

\subsection{Geometric algebra}

Geometric calculus (GC) is the calculus formalism based on GA. The GA of an $n$-dimensional space $\C{n}$ is motivated by the desire to find a closed algebra on a vector space with respect to multiplication\footnote{A formal definition of Geometric Algebra and the geometric product can be found in \cite{Macdonald_2002}}. We obtain it when we equip a vector space with a product over its elements $ u, v, w \in \mathcal{V} $ that is:
\begin{itemize}
	\item Associative, $ (uv)w = u(vw) $
	\item Left-distributive, $ u (v + w) = uv + uw $
	\item Right-distributive, $ (v + w)u = vu + wu $
	\item Reduces to the usual inner product, $ u^2 = g(u,u) $.
\end{itemize}

These properties are satisfied by the \textit{geometric product}\cite{Hestenes2003a_OerstedMedal}, which we can write as the sum of the inner and outer product of vectors,
\begin{equation}\label{eq:Geometric_Product}
	uv := u \cdot v + u \wedge v.
\end{equation}

The geometric product is, in general, not symmetric nor anti-symmetric. If the vectors are parallel, it is symmetric. If they are orthogonal, it is anti-symmetric. Therefore, we must take special care in keeping the appropriate order of operations. Nonetheless, it is associative.

$ \C{n} $ of a vector space $\mathcal{V}$ of dimension $n$ is composed by the direct sum of the exterior algebras of dimension $k$, with $k$ going from $0$ to $n$,
\begin{equation}
	\C{n} = \bigoplus_{k=0}^n \bigwedge^k\left(\mathcal{V}\right).
\end{equation}

Therefore, $\C{n}$ has $ \sum_{k=0}^{n} \binom{n}{k} = 2^n$ dimensions. We call the elements of $\bigwedge^k$ $k$-vectors, and we say they have \textit{grade} $k$, corresponding to their exterior power in differential forms. Each $k$-vector of $ \C{n} $ represents an oriented geometric object of the space. An object composed of multiple $k$-vectors of different grade is called a \textit{multivector}, while an object composed solely by the outer product of vectors is called a \textit{blade}\footnote{Only in spaces of dimension 4 or greater we can construct single-grade elements that are not blades. e.g.: $e_1e_2 + e_3e_4$}.

For example, in the Euclidean plane, $ \mathbb{E}^2 $, with base elements $ \{e_1,e_2\} $, we can construct the geometric algebra $ \C{2} $, with base elements $ \{1, e_1, e_2, e_1\wedge e_2\} $. Where $ 1 $ represents points (or scalars), $ \{e_1, e_2\} $ represent vectors and $ e_{1} \wedge e_2 $ represents an oriented plane. All the $k$-vectors of a space can be encoded in a single \textit{multivector}, which in the case of $ \C{2} $ would be
\begin{equation} \label{eq:Multivector_E2}
	M = M^0 + M^1e_1 + M^2 e_2 + M^3 e_1 \wedge e_2,
\end{equation}
being $ \{M^i\}_{i=0,1,2,3} \in \mathbb{R}$ scalars.

The element of the highest grade of a space is called pseudoscalar; it is designed by $I$, and it has the same dimensionality as the vector space $\mathcal{V}$. The pseudoscalar holds special importance in a GA because it allows us to define the \textit{duality operation}, which geometrically maps any space element with grade $m$ to an element of grade $n-m$. For example, let's take a bivector in $\mathbb{E}^3$, $M = M^{12} e_1 \wedge e_2$, its dual would be
\begin{equation}
	I M = e_{1} \wedge e_2 \wedge e_3 M^{12}  e_{1} \wedge e_2 =  -M^{12} e_3.
\end{equation}
Where we have used the anticommutativity of orthogonal basis vectors to change their order and the associativity of the geometric product to simplify the expression $(e_1 \wedge e_2 \wedge e_3)(e_1 \wedge e_2) = e_1 e_2 e_3 e_1 e_2 = (e_1 e_1) e_2 e_3 e_2 = - (e_2 e_2) e_3 = -e_3$. Note that the action the left and right action of $I$ are in general not equivalent.

The dual operation in GA is mathematically equivalent to the Hodge dual in differential forms and creates a direct correspondence between the outer product and the cross product in three dimensions
\begin{equation}
	a \times b = -(a \wedge b)I.
\end{equation}
This correspondence only holds in three dimensions because only in three dimensions the dual of a bivector is a vector. That is the main reason why the outer product is a preferable to the cross product. With the outer product we can to define rotations by the plane of rotation instead of its axis and efficiently perform rotations in higher-than-two dimension spaces.

An essential feature of the geometric product, which is not shared with the inner or outer product, is that it can be invertible. For a non-null vector $a$, we can define $a^{-1} = a /a^2$ such that $a a^{-1} = 1$. The reason being its associativity and the existence of elements of mixed grade, which allows \cref{eq:Geometric_Product} to contain all the geometric information of the relative geometric position of $a$ and $b$.

In order to simplify notation by saving unnecessary parenthesis, we will adopt the standard convention of always performing inner and outer products before geometric products.

\subsection{Complex numbers and rotor techniques}

The use of the geometric product as the cornerstone of GA may seem unimportant, but it is the fundamental source of its notational power. Two examples of this are its treatment complex numbers and rotations. 

Starting with $\C{2}$, let's calculate the square of the bivector element
\begin{equation}
	(e_1 \wedge e_2)^2 = e_1 e_2 e_1 e_2 = - (e_1 e_1) (e_2 e_2) = -1.
\end{equation}

Because a multivector can have multiple grades, we can construct an element of the even sub algebra of such a space, called $\C{2}^+$, in the form
\begin{equation}
	z = \alpha + \beta e_1e_2.
\end{equation}
With $\alpha,\beta \in \mathbb{R}$. It's self-evident that $z$ is isomorphic to the complex number $\alpha + \beta i$.

Moreover, we can use the bivector $e_1e_2$ as a generator of rotations in the $e_1e_2$- plane. As a simple example we can see that left multiplication by $e_1e_2$ indeed produces the $\pi/2$ rotations expected from a complex unit
\begin{equation}
	e_1 (e_1e_2) = e_2, \quad e_2 (e_1e_2) = -e_1.
\end{equation}
Furthermore, we can use the bivector to construct a rotor
\begin{equation}
	R(\theta) = \exp\left(-\hf \theta  e_1e_2\right) = \cos\frac{\theta}{2} - e_1e_2\sin\frac{\theta}{2},
\end{equation}
which generates a rotation of an angle $\theta$ in the $e_1e_2$-plane of any multivector $M$ when acting as
\begin{equation}
	M' = R(\theta) M R^\dagger(\theta).
\end{equation}
With $R^\dagger(\theta) = \exp\left(- \hf \theta e_2e_1\right) = \exp\left(\hf \theta  e_1e_2\right)$, being \( R(\theta) \) with the bivector order interchanged.

These expressions are valid not only in $\mathbb{E}^2$, but in spaces of any dimension and signature. Providing a direct generalization of complex numbers. As a simple example, the bivectors of $\C{3}$ are isomorphic to the quaternions
\begin{equation}
	e_1 \wedge e_2 \leftrightarrow i,\quad e_2 \wedge e_3\leftrightarrow j, \quad e_3 \wedge e_1 \leftrightarrow -k.
\end{equation}

Besides the mathematical equivalence of complex numbers and the even subalgebras of GA, it should be noted that GA provides a geometric meaning that imaginary numbers lack. For example, by observing the exponent of a plane wave in GA, one immediately knows the polarization plane(s), and in case of having multiple waves colliding, the formalism automatically takes care of how the different polarization planes combine.

These features are direct consequence of having mixed grade elements are characteristic of GA, and are not translatable to differential forms language. 

\subsection{The vector derivative}

Geometric Calculus(GC) comprises the differential and integral techniques developed from GA. The main derivative operator in GC is the vector derivative $\nabla$. It can be written in a particular base as
\begin{equation}
	\nabla = \sum_{i=1}^{n} e^i \p_i.
\end{equation}
Where $e^i $ is the reciprocal base of $e_i$, defined by the relation $e^i e_j = \delta^i_j$, and $\p_i$ is the usual directional derivative in the $i$th-direction of a scalar function $f(x)$
\begin{equation}
	\p_i f(x) = \frac{\p}{\p x^i} f(x) = \lim_{\epsilon \to 0} \frac{f(x+\epsilon e_i)-f(x)}{\epsilon}.
\end{equation}

$\nabla$ has all the algebraic properties of a vector and can act over any multivector with the inner, outer or geometric product, producing all vector differential operators.

Acting on a scalar field $ \phi = \phi(x) $, the vector derivative produces a gradient,
\begin{equation}
	\nabla \phi (x) = e^i \p_i \phi(x).
\end{equation}

And acting on a vector field $ v = v(x) = v^i(x) e_i = v^i e_i$ produces
\begin{itemize}
	\item Geometric product: $ \nabla v = \nabla \cdot v + \nabla \wedge v $
	\item Divergence: $  \nabla \cdot v = \p_i v^i$
	\item ``Curl" - Exterior derivative:  $ \nabla \wedge v =  \sum_{i, j=0}^{3} \left(\p_i v^j\right) e^i \wedge e_j$
	\item Laplacian: $ \nabla^2 v = \nabla (\nabla v) = (\nabla \nabla) v= \sum_{i=0}^{3} \p_i^2 v$
\end{itemize}

The vector derivative behaves algebraically as a vector, and therefore, it follows the same rules as any object of grade 1. For example, being $\phi $ a scalar field and $a$ and $b$ two vector fields.
\begin{itemize}
	\item $\nabla (\phi a ) = (\nabla \phi ) a + \phi (\nabla a)$
	\item $\nabla (a \wedge b) = (\nabla a) \wedge b - a \wedge (\nabla b)$
\end{itemize}
Its commutativity or anticommutativity depends on the object that it is acting upon. A good summary of its properties when acting over different elements can be found in \cite{Hitzer2013}.

It follows from the symmetry properties of $\nabla$ that for a $k$-vector field $a = \sum_{i}a_{(i)}^{1...k} e_{1} \wedge ... \wedge e_{k}$
\begin{equation}\label{eq:Double_wedge}
	\nabla \wedge \nabla \wedge a = 0.
\end{equation}

The proof is straightforward.
\begin{equation}
	\nabla \wedge a = \sum_{j, i} \p_j a_{(i)}^{1...k} e^j \wedge e_{1} \wedge ... \wedge e_{k}
\end{equation}
\begin{equation}
	\nabla \wedge \nabla \wedge a = \sum_{l, j, i} \p_l \p_j a_{(i)}^{1...k} e^l \wedge e^j \wedge e_{1} \wedge ... \wedge e_{k} = 0
\end{equation}
By the symmetry of the partial derivatives $\p_l \p_j$ contracted with the anti-symmetry of $ e^l \wedge e^j$.

Finally, we should mention that $\nabla$ is an invertible operator thanks to the geometric product. As we will see in \cref{sec:Maxwell}, this allows us to directly solve some first-order differential equations without passing through second-order equations.

\subsection{Fundamental theorem of calculus}\label{sec:fundamental-theorem-of-calculus}

Integration in GC has two main benefits compared to vector calculus: directed integration and the geometric product. The first is a feature shared with differential forms. Exterior algebra allows considerable simplification when dealing with geometrical, oriented elements. The second advantage is characteristic of GC. The vector character of $\nabla$ and the properties of the geometric product expand the applicability of the integral theorems of differential forms and, in some cases, allow for more straightforward computations.

\begin{figure}
	\centering
	\def\svgwidth{0.7\textwidth}
\begingroup%
  \makeatletter%
  \providecommand\color[2][]{%
    \errmessage{(Inkscape) Color is used for the text in Inkscape, but the package 'color.sty' is not loaded}%
    \renewcommand\color[2][]{}%
  }%
  \providecommand\transparent[1]{%
    \errmessage{(Inkscape) Transparency is used (non-zero) for the text in Inkscape, but the package 'transparent.sty' is not loaded}%
    \renewcommand\transparent[1]{}%
  }%
  \providecommand\rotatebox[2]{#2}%
  \newcommand*\fsize{\dimexpr\f@size pt\relax}%
  \newcommand*\lineheight[1]{\fontsize{\fsize}{#1\fsize}\selectfont}%
  \ifx\svgwidth\undefined%
    \setlength{\unitlength}{269.94589017bp}%
    \ifx\svgscale\undefined%
      \relax%
    \else%
      \setlength{\unitlength}{\unitlength * \real{\svgscale}}%
    \fi%
  \else%
    \setlength{\unitlength}{\svgwidth}%
  \fi%
  \global\let\svgwidth\undefined%
  \global\let\svgscale\undefined%
  \makeatother%
  \begin{picture}(1,0.26515291)%
    \lineheight{1}%
    \setlength\tabcolsep{0pt}%
    \put(0,0){\includegraphics[width=\unitlength,page=1]{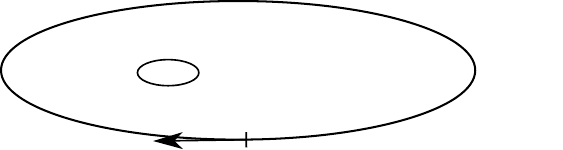}}%
    \put(0.49535907,0.12449834){\makebox(0,0)[lt]{\lineheight{1.25}\smash{\begin{tabular}[t]{l}$\Omega$\end{tabular}}}}%
    \put(0.79162558,0.21312216){\makebox(0,0)[lt]{\lineheight{1.25}\smash{\begin{tabular}[t]{l}$\partial \Omega$\end{tabular}}}}%
    \put(0.23650488,0.17544863){\makebox(0,0)[lt]{\lineheight{1.25}\smash{\begin{tabular}[t]{l}$\mathrm{d} x_p$\end{tabular}}}}%
    \put(0.33265119,0.03937832){\makebox(0,0)[lt]{\lineheight{1.25}\smash{\begin{tabular}[t]{l}$\mathrm{d} x_{p-1}$\end{tabular}}}}%
    \put(0,0){\includegraphics[width=\unitlength,page=2]{Integration.pdf}}%
  \end{picture}%
\endgroup%

	\caption{Depiction of the a $p$-dimensional surface $\Omega$ with differential oriented element $\dd x_p$, bounded by the $(p-1)$-dimensional surface $\p \Omega$ with differential oriented element $d x_{p-1}.$} 
	\label{fig:Integration}
\end{figure}

The fundamental theorem of geometric calculus is stated as follows \cite{Macdonald_GC,Hestenes_STA}: \textit{Consider a closed $p$-dimensional surface, $\Omega$, bounded by the $(p-1)$-dimensional surface $\p \Omega$. We can represent a differential oriented element of $\Omega$ by the $p$-vector $\dd x_p$, and the differential oriented element of the boundary by $\dd x_{p-1}$, \cref{fig:Integration}. If $M$ is a differentiable multivector function on $\Omega$, then we can write}
\begin{equation}\label{eq:Fundamental_Theorem}
	\int_{\Omega} \dd x_p \cdot \nabla M = \oint_{\partial \Omega} \dd x_{p-1} M
\end{equation}

Notice that $\dd x_p$ can be decomposed as the outer product of differential vectors $ \dd x^{(i)} = \dd x^i e_i$ (no summing convention),
\begin{equation}
	\dd x_p = \dd x^{(1)} \wedge ... \wedge \dd x^{(p)} = \dd x^1 ... \dd x^p e_1 \wedge...\wedge e_p.
\end{equation}
$e_1 \wedge...\wedge e_p$ is the unitary pseudoscalar of the surface $\Omega$, and $\dd x^i$ are the scalar integration elements. We can perform an analogous decomposition for $\dd x_{p-1}$.

In \cref{eq:Fundamental_Theorem}, $\nabla = \sum_{i=1}^{p} e^i \p_i $ is the vector derivative in $\Omega$, and it is important to remark that $ \dd x_p \wedge \nabla = 0 $ because $\dd x_p$ has the same dimension as $\Omega$, and therefore $ \dd x_p \nabla = \dd x_p \cdot \nabla $. The inner product with the vector $\nabla$ lowers by one the grade of $\dd x_p$, making $\dd x_p \cdot \nabla$ of the same grade as $ \dd x_{p-1} $, which is essential if \cref{eq:Fundamental_Theorem} is to make sense.

The fundamental theorem of geometric calculus applies to manifolds of any dimensions and admits the presence of holes. The boundaries facing the holes are part of $\p \Omega$ \cite{Macdonald_GC}.

The main differences with the generalized Stokes' theorem of differential forms are:
\begin{itemize}
	\item In \cref{eq:Fundamental_Theorem}, the product between $\dd x_p \cdot \nabla$ and $M$ is a geometric product. Therefore, it decomposes in different parts corresponding to the generalized Stokes' theorem and its dual.
	\item $M$ is not limited to being a map to scalars as in differential forms. It is a general map from multivectors to multivectors. This allows \cref{eq:Fundamental_Theorem} to incorporate Cauchy's integral theorem.	
\end{itemize}

The simplicity of \cref{eq:Fundamental_Theorem} can hinder its power. So, as an illustration, we will apply it to different manifolds to reproduce the fundamental theorem of calculus, the Divergence theorem, Curl theorem, Green's theorem and Cauchy's integral theorems of complex analysis as they are shown in \cite{Macdonald_GC}.

\paragraph{Fundamental theorem of calculus} If $p = 1$, $\Omega$ is a line and $\dd \omega \rightarrow \dd \vec{s}$ is a vector. $M$ is a scalar function, $\p \Omega$ are the starting and end points of the segment, and integration over them consists of multiplying them by $1$ or $-1$, depending on the orientation of the segment. Therefore, \cref{eq:Fundamental_Theorem} reduces to the usual fundamental theorem of calculus

\begin{equation}
	\int_{x_1}^{x_2} \vec{\dd s} \cdot M = M(x_2)- M(x_1)
\end{equation}

\paragraph{Divergence theorem} Let $M$ be a vector field $v$ on a bounded $p$-dimensional manifold $\Omega$ in $\mathbb{R}^p$. Setting $\dd\sigma = \vec{n} \dd_{p-1}x$, \cref{eq:Fundamental_Theorem} can be written as
\begin{equation}
	\int_\Omega \nabla \cdot v \:\dd_p x = \oint_{\p\Omega} v\cdot \dd \sigma
\end{equation}
If we specify it for a 3-dimensional space, the Divergence theorem is called \textit{Gauss' theorem}. However, this covers only the inner product part of \cref{eq:Fundamental_Theorem}. The outer product part, together with the dual operation, relates outer and cross products in $\mathbb{R}^3$ and produces
\begin{equation}
	\iiint_V \nabla \times v \:\dd V = \oiint_S \dd \sigma \times v.
\end{equation}

\paragraph{Curl theorem} Let $M$ be a $(p-1)$-vector valued field on an oriented and bounded $p$-dimensional manifold $\Omega \in \mathbb{R}^n, p \leq n$. Then, \cref{eq:Fundamental_Theorem} is
\begin{equation}
	\int_\Omega \dd_p x \cdot (\nabla_p \wedge M) = \oint_{\p\Omega} \dd_{p-1}x \cdot M.
\end{equation}
If we specify it for a 3-dimensional manifold, with a 2-dimensional surface $\Omega$ with a differential of area $d\sigma$, $\p\Omega$ as the enclosing curve with a differential of the line $ds$ and being $M$ a vector field $v$. Then, Curl's theorem is known as \textit{Stokes' theorem},
\begin{equation}
	\iint_S \left(\nabla \times v\right) \cdot \dd \sigma = \oint_C v \cdot \dd s.
\end{equation}

\subparagraph{Green's theorem} Let $R$ be a region of the $xy$-plane with boundary $C$ oriented counterclockwise. Let $M$ be a vector field $P(x,y)\hat{x} + Q(x,y)\hat{y}$ on $R$. Then
\begin{equation}
	\iint_R \left(\p_x Q - \p_y P\right)\dd A = \oint_C (P\dd x + Q \dd y)
\end{equation} 

\paragraph{Cauchy's theorem} The concept of analytic functions in the complex plane can be generalized by geometric calculus to multivector fields $M$ that satisfy the condition
\begin{equation}
	\nabla M = 0.
\end{equation}
This generalization effectively reflects the idea that holomorphic functions in complex analysis correspond to exact differential forms. For those analytic functions, \cref{eq:Fundamental_Theorem} simplifies to
\begin{equation}
	\oint_{\partial \Omega} \dd x_{p-1} M = 0,
\end{equation}
which is called the \textit{Generalized Cauchy's theorem}.

Geometric calculus also includes the \textit{residue theorem}: If $M$ is analytic except at a pole $x'$ with residue $q$ in the 2-dimensional surface $\mathcal{A}$, then $\nabla M = 2\pi q \delta(x-x')$, and substitution into \cref{eq:Fundamental_Theorem} produces \cite{Hestenes_1986}
\begin{equation}
	\oint \dd x M = 2\pi q \int_{\mathcal{A}} e_1\wedge e_2 |\dd A(x)| \delta(x-x') = 2\pi e_1\wedge e_2 q
\end{equation}
Notice how $e_1\wedge e_2$ plays the role of the imaginary unit $i$.

Details for the derivation of Cauchy's theorem for integral functions can be found in \cite{Dressel2015, Doran2013, Hestenes_GC}, which we will not reproduce here. Nonetheless, we would like to highlight three points from the result:
\begin{itemize}
	\item The theorem is derived without introducing complex numbers because the complex plane is isomorphic to $\mathbb{C}^2$.
	\item The result applies to spaces of arbitrary dimensions.
	\item The theorem has a clear geometrical interpretation.
	\item GC unifies the theory of poles and residues with that of Green's and delta functions.
\end{itemize}

The unification power of the fundamental theorem of geometric calculus is unique to GC and it provides a clear geometrical interpretation of most integral theorems. For more details on geometric calculus or the fundamental theorem of calculus, we recommend the accessible book \cite{Macdonald_GC} or the more technical \cite{Hestenes_GC}.

\section{Differential forms in Geometric Calculus}

\subsection{Differential forms}\label{sec:differential-forms}

The correspondence between differential forms and geometric calculus can be confusing because differential forms are used in literature in two distinct ways: As a basis of the cotangent space and as differential elements for integration.

The first case is equivalent to using partial derivatives as basis elements of the tangent space, so let us discuss that first. This identification has a long tradition in mathematics and it is based on existence of an isomorphism between coordinate basis vectors and partial derivatives. It is undoubtedly true that this identification works and leads to an efficient way of describing differential geometry.

However, in the opinion of the authors, the link between flat and curved manifolds is reduced and a better geometric meaning is retained by not performing this identification. As we will see in this section, GC allows for a description that is equal, if not more, powerful and compact than the one provided by differential forms while providing an easier geometric interpretation.

\begin{figure}
	\centering
	\def\svgwidth{0.7\textwidth}
	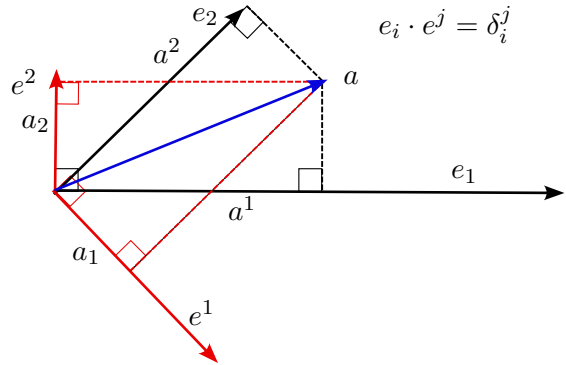
	\caption{Normal and reciprocal basis of $\mathbb{R}^2$, depicted in black and red respectively. Both basis are related by an orthonormality condition $e_i \cdot e^j = \delta_i^j$. A vector $a$ is depicted in blue with components $a^i$ in the $\{e_i\}$ basis and components $a_i$ in the reciprocal $\{e^i\}$ basis. Notice that both basis cover the same physical space.} 
	\label{fig:reciprocal-basis}
\end{figure}

Therefore in GC we will denote the basis vectors of the (tangent) space as $e_i$ and the basis vectors of the reciprocal\footnote{We will use the word reciprocal instead of cotangent, in line with its the use in solid state physics, to emphasize that the normal and reciprocal basis cover the same physical space.} space as $e^i$, which relate by the usual relation
\begin{equation}
	e_i \cdot e^j = \delta_i^j.
\end{equation}
The key difference between GC and differential forms is that in GC both basis are vector basis of \textit{the same physical space}, see \cref{fig:reciprocal-basis}.

This construction of course is only possible in metric spaces, where the musical isomorphism relates the tangent and cotangent spaces. However, the presence of a metric is a requirement to construct the geometric product and a GA. Because we are focusing in physical applications, where a metric is always present, we will not discuss any case where we cannot perform this identification.

When differential forms are used as basis of the space, they can be identified with unitarian basis elements of the space $\C{n}$,
\begin{itemize}
	\item zero-form $\leftrightarrow$ Scalar.
	\item one-form $\leftrightarrow$ Vector $\hat{e}_i$
	\item two-form $\leftrightarrow$ Bivector $A = \hat{e}_1 \wedge \hat{e}_2$.
\end{itemize}

In the second use of differential forms, as directed integration elements we identify them with differential multivectors representing differential oriented geometric elements,
\begin{itemize}
	\item zero-form $\leftrightarrow$ Scalar.
	\item one-form $\leftrightarrow$ Differential vector $\dd x = |\dd x^i| \hat{e}_i$
	\item two-form $\leftrightarrow$ Differential of area $\dd A = \dd x^{(1)} \wedge \dd x^{(2)} = |\dd x^1||\dd x^2| \hat{e}_1 \wedge \hat{e}_2$.
\end{itemize}
Where $|\dd x^i|$ represents the (differential) norm of the (differential) vector $\dd x^i$, corresponding to the usual scalar element of integration, and $\hat{e}_i$ are unitary basis elements. E.g., In polar coordinates, the differential of the area would be
\begin{equation}
	\dd x^r \wedge \dd x^\phi = |\dd x^r||\dd x^\phi| e_r \wedge e_\phi = \: \dd r \: \dd\phi \: \hat{e}_r \wedge r \hat{e}_\phi = r \: \dd r \: \dd \phi \:\hat{e}_r \wedge \hat{e}_\phi.
\end{equation}
Recovering the polar element of area times the unitary bivector $\hat{e}_r \wedge \hat{e}_\phi$, expressing the orientation of the integral.

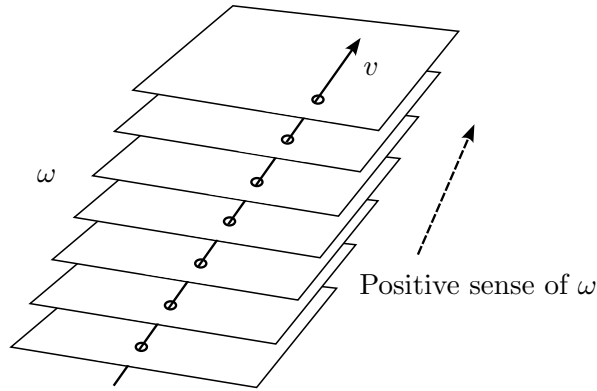
\begin{figure}[h!]
	\centering
	\def\svgwidth{0.5\textwidth}
\begingroup%
  \makeatletter%
  \providecommand\color[2][]{%
    \errmessage{(Inkscape) Color is used for the text in Inkscape, but the package 'color.sty' is not loaded}%
    \renewcommand\color[2][]{}%
  }%
  \providecommand\transparent[1]{%
    \errmessage{(Inkscape) Transparency is used (non-zero) for the text in Inkscape, but the package 'transparent.sty' is not loaded}%
    \renewcommand\transparent[1]{}%
  }%
  \providecommand\rotatebox[2]{#2}%
  \newcommand*\fsize{\dimexpr\f@size pt\relax}%
  \newcommand*\lineheight[1]{\fontsize{\fsize}{#1\fsize}\selectfont}%
  \ifx\svgwidth\undefined%
    \setlength{\unitlength}{457.02401733bp}%
    \ifx\svgscale\undefined%
      \relax%
    \else%
      \setlength{\unitlength}{\unitlength * \real{\svgscale}}%
    \fi%
  \else%
    \setlength{\unitlength}{\svgwidth}%
  \fi%
  \global\let\svgwidth\undefined%
  \global\let\svgscale\undefined%
  \makeatother%
  \begin{picture}(1,0.8032877)%
    \lineheight{1}%
    \setlength\tabcolsep{0pt}%
    \put(0,0){\includegraphics[width=\unitlength,page=1]{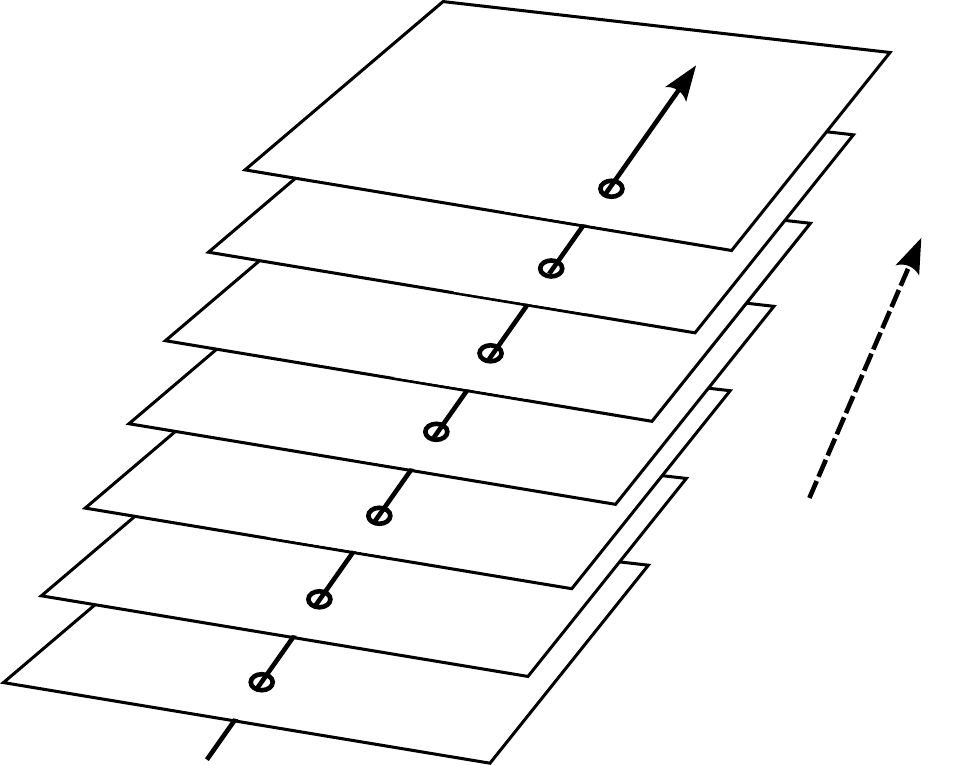}}%
    \put(0.72827324,0.1970604){\color[rgb]{0,0,0}\makebox(0,0)[lt]{\lineheight{1.25}\smash{\begin{tabular}[t]{l}Positive sense of $\omega$\end{tabular}}}}%
    \put(0.73718584,0.65754513){\color[rgb]{0,0,0}\makebox(0,0)[lt]{\lineheight{1.25}\smash{\begin{tabular}[t]{l}$v$\end{tabular}}}}%
    \put(0.05834221,0.43175905){\color[rgb]{0,0,0}\makebox(0,0)[lt]{\lineheight{1.25}\smash{\begin{tabular}[t]{l}$\omega$\end{tabular}}}}%
  \end{picture}%
\endgroup%

	\caption{Depiction of a one-form $\omega$ pierced by the vector $v$. The resulting scalar, $\omega(v)$ is ``equivalent" to the number of surfaces that $v$ goes through.} 
	\label{fig:One-form}
\end{figure}

\paragraph{Geometric interpretation} Differential one-forms are typically depicted as surfaces that are pierced by vectors, \cref{fig:One-form}, while two-forms are interpreted as the ``tubes" that result from the intersections of two one-forms \cite{Misner1973}. Despite the arguments made in the literature about the intuitiveness of this description, this description is quite difficult to imagine and much more to attribute a physical interpretation to it. Higher dimensional elements get more and more difficult to imagine.

Of course, with enough practice, one can learn to think about forms as planes, their intersections and how they are pierced by vectors, but only after a specific explanation has been given and some training has been performed.

\begin{figure}
	\centering
	\includegraphics[width=0.9\textwidth]{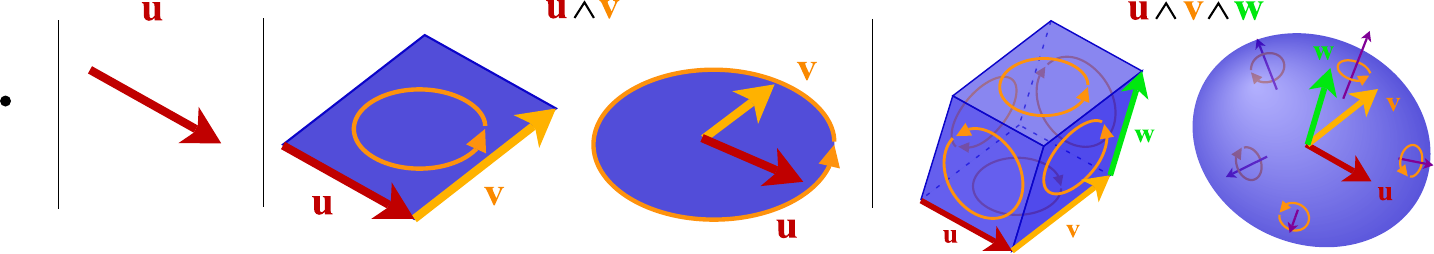}
	\caption{Geometrical elements of $\C{3}$. From top to bottom, they correspond to scalars, vectors, bivectors and trivectors(volumes). Notice that $k$-vectors only contain information about the ``area/volume" and its orientation not about the shape.} 
	\label{fig:Exterior-algebra}
\end{figure}

On the other hand, geometric calculus provides a straightforward geometrical interpretation of quantities. Vectors are directed lines; bivectors are directed planes; trivectors are oriented volumes, \cref{fig:Exterior-algebra}. Their outer product is the direct geometrical combination between elements, and their inner product is the projection onto each other. Based on our experience teaching geometric calculus, its exposition is way more intuitive than forms. Usually, students who have received minimal exposure to geometric algebra can immediately explain the geometrical meaning of different elements and operations between them. This is rarely the case with differential forms and their operations.

\begin{figure}[h!]
	\centering
	\def\svgwidth{0.4\textwidth}
\begingroup%
  \makeatletter%
  \providecommand\color[2][]{%
    \errmessage{(Inkscape) Color is used for the text in Inkscape, but the package 'color.sty' is not loaded}%
    \renewcommand\color[2][]{}%
  }%
  \providecommand\transparent[1]{%
    \errmessage{(Inkscape) Transparency is used (non-zero) for the text in Inkscape, but the package 'transparent.sty' is not loaded}%
    \renewcommand\transparent[1]{}%
  }%
  \providecommand\rotatebox[2]{#2}%
  \newcommand*\fsize{\dimexpr\f@size pt\relax}%
  \newcommand*\lineheight[1]{\fontsize{\fsize}{#1\fsize}\selectfont}%
  \ifx\svgwidth\undefined%
    \setlength{\unitlength}{184.68212218bp}%
    \ifx\svgscale\undefined%
      \relax%
    \else%
      \setlength{\unitlength}{\unitlength * \real{\svgscale}}%
    \fi%
  \else%
    \setlength{\unitlength}{\svgwidth}%
  \fi%
  \global\let\svgwidth\undefined%
  \global\let\svgscale\undefined%
  \makeatother%
  \begin{picture}(1,0.81325875)%
    \lineheight{1}%
    \setlength\tabcolsep{0pt}%
    \put(0,0){\includegraphics[width=\unitlength,page=1]{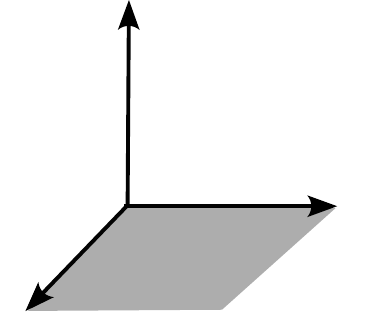}}%
    \put(0.00237697,0.14409176){\color[rgb]{0,0,0}\makebox(0,0)[lt]{\lineheight{1.25}\smash{\begin{tabular}[t]{l}$e_1$\end{tabular}}}}%
    \put(0.84536655,0.3214282){\color[rgb]{0,0,0}\makebox(0,0)[lt]{\lineheight{1.25}\smash{\begin{tabular}[t]{l}$e_2$\end{tabular}}}}%
    \put(0.38341905,0.75264476){\color[rgb]{0,0,0}\makebox(0,0)[lt]{\lineheight{1.25}\smash{\begin{tabular}[t]{l}$e_3$\end{tabular}}}}%
    \put(0,0){\includegraphics[width=\unitlength,page=2]{vector-projection.pdf}}%
    \put(-0.00215043,0.76107528){\color[rgb]{0,0,0}\makebox(0,0)[lt]{\lineheight{1.25}\smash{\begin{tabular}[t]{l}$a$\end{tabular}}}}%
    \put(0.55140092,0.31391752){\color[rgb]{0,0,0}\makebox(0,0)[lt]{\lineheight{1.25}\smash{\begin{tabular}[t]{l}$a \cdot B$\end{tabular}}}}%
    \put(0,0){\includegraphics[width=\unitlength,page=3]{vector-projection.pdf}}%
    \put(0.41276467,0.12261058){\color[rgb]{0,0,0}\makebox(0,0)[lt]{\lineheight{1.25}\smash{\begin{tabular}[t]{l}$B$\end{tabular}}}}%
  \end{picture}%
\endgroup%

	\caption{Inner product of the vector $a = \hat{e}_1 + \hat{e}_3$ with the bivector $B = \hat{e}_1 \wedge \hat{e}_2$.} 
	\label{fig:vector-projection}
\end{figure}

Let's take as an example the inner product of a vector $a = \hat{e}_1 + \hat{e}_3$ with the bivector $B = \hat{e}_1\wedge \hat{e}_2$, \cref{fig:vector-projection}
\begin{equation}
	a \cdot B = (\hat{e}_1 + \hat{e}_3)\cdot \hat{e}_1\wedge \hat{e}_2 = \hat{e}_2.
\end{equation}
The geometric construction is very simple, project $a$ onto the plane defined by $B$, and rotate it $\pi/2$ in the direction defined by $B$. The rotation might come as a surprise, but one should remember that bivectors are also the generators of rotations. Note that the resulting vector is contained in the plane defined by $B$ and it is always orthogonal to $a$.

The equivalent operation in differential forms is not even possible to imagine, since the piercing picture of one-forms and vectors only works when for obtaining scalar quantities, represented by the ``amount" of surfaces the vector goes through.

\subsection{Operations}

\paragraph{Duality operation} We perform the duality operation in differential forms with the Hodge star operator. In geometric calculus, we obtain the dual of an object by multiplying by the pseudoscalar $I$. Sometimes, we should add a minus sign, depending on the grade of the objects and the side of the $I$ multiplication.

Taking $\mathbb{E}^3$ as an example,
\begin{itemize}
	\item Dual of a zero form $\star f = f \dd x^1 \wedge \dd x^2 \wedge \dd x^ 3 \leftrightarrow I f = f \hat{e}^1 \wedge \hat{e}^2 \wedge \hat{e}^3$.
	\item Dual of a one-form $\star \dd x^1 = \dd x^2 \wedge \dd x^3 \leftrightarrow I \hat{e}^1 = \hat{e}^1 I=  \hat{e}^2 \wedge \hat{e}^3 $.
	\item Dual of a two-form $\star (\dd x^1\wedge \dd x^2) = \dd x^3 \leftrightarrow I \hat{e}^1 \wedge \hat{e}^2 = \hat{e}^1 \wedge \hat{e}^2 I = - \hat{e}^3 $.
\end{itemize}

The commutativity of the pseudoscalar with a $k$-vector $A_k$ in an $n$-dimensional space is given by 
\begin{equation}
	I A_k = (-1)^{k(n-k)}A_k I.
\end{equation}
The minus sign depends on the number of permutations we must perform to bring the pseudoscalar to the other side. Thus, the pseudoscalar commutes with all elements in spaces of odd dimension. While in spaces of even dimensions, the pseudo scalar commutes with even elements and anti-commutes with odd elements.

\paragraph{Outer product} The outer product of differential forms is equivalent to the outer product in geometric calculus. Both are superior to the cross-product of conventional vector calculus. The outer product allows us to perform consistently rotations and curls in spaces of any dimension greater than 1.

Being $\sigma, \omega$ one-forms and $a, b$ vectors, their outer product is
\begin{equation}
	\begin{aligned}
		\sigma \wedge \omega & = (\sigma_1 \omega_2 - \sigma_2 \omega_1)\dd x^1 \wedge \dd x^2\\
		& + (\sigma_2 \omega_3 - \sigma_3 \omega_2)\dd x^2 \wedge \dd x^3\\
		& + (\sigma_3 \omega_1 - \sigma_1 \omega_3)\dd x^3 \wedge \dd x^1 \\
		& \Updownarrow\\
		a \wedge b & = (a_1 b_2 - a_2 b_1) \hat{e}^1 \wedge \hat{e}^2\\
		& + (a_2 b_3 - a_3 b_2) \hat{e}^2 \wedge \hat{e}^3\\
		& + (a_3 b_1 - a_1 b_3) \hat{e}^3 \wedge \hat{e}^1
	\end{aligned}
\end{equation}

They both relate to the cross-product of vector calculus using the duality operation. In three dimensions,
\begin{equation}
	a \times b \Leftrightarrow \star (a \wedge b) \Leftrightarrow -I(a \wedge b).
\end{equation}

\paragraph{Inner product} 

The inner product in differential forms is identical to the one in geometric algebra\footnote{In older texts on differential forms, there was no product and instead the equivalent expression $\star (\sigma \wedge \star \omega)$ was used.\label{foot:old-inner-product}}
\begin{equation}
	\begin{aligned}
		\langle\sigma,\omega\rangle & = \sigma_1\omega_1 + \sigma_2\omega_2 + \sigma_3 \omega_3\\
		& \Updownarrow\\
		a \cdot b &= a_1 b_1 + a_2 b_2 + a_3 b_3
	\end{aligned}
\end{equation}

In the differential forms treatment it is sometimes important to differentiate between the inner and the interior product. The interior product is defined as the action of a one-form over a vector and it is metric independent, while the inner product is the usual metric product between two vectors. In case of having a metric space, there is an isomorphism between vectors and one-forms and both products are equivalent.

From the GA perspective, where difference between inner and interior products simply lies in which frame you are expressing your vectors, $a = a^i e_i = a_i e^i$,
\begin{equation}
	a \cdot b = a^i b^j (e_i \cdot e_j) = a^i b^j g_{ij} = a_i b^j (e^i \cdot e_j) = a_i b^j \delta^i_j = a^i b_i.
\end{equation}

\subsubsection{Differential operations}

Because the vector derivative can be treated algebraically as a vector, all algebraic identities of relating vectors, inner, outer and geometric products apply\cite{Hitzer2013}, generating the different differential identities. Here, we will present some of them and see how they relate to the usual operations in differential forms.

\paragraph{Gradient:} The gradient of a scalar function, or zero form, corresponds to its exterior derivative and to its curl in geometric calculus
\begin{equation}
	\begin{aligned}
		\mathbf{d}f & = \p_1 f \dd x^1 + \p_2 f \dd x^2 + \p_3 f \dd x^3\\
		& \Updownarrow\\
			\nabla f & = \nabla \wedge f = \left(\p_1 f\right) \hat{e}^1 + \left(\p_2 f\right) \hat{e}^2 + \left(\p_3 f\right) \hat{e}^3\\
	\end{aligned}
\end{equation}

\paragraph{Exterior derivative:} The exterior derivative in differential forms corresponds to the curl of a $k$-vector in geometric calculus,
\begin{equation}
	\begin{aligned}
		\mathbf{d} \omega & = (\p_1 \omega_2 - \p_2 \omega_1)\dd x^1 \wedge \dd x^2\\
		& + (\p_2 \omega_3 - \p_3 \omega_2)\dd x^2 \wedge \dd x^3\\
		& + (\p_3 \omega_1 - \p_1 \omega_3)\dd x^3 \wedge \dd x^1 \\
		& \Updownarrow\\
		\nabla \wedge a & = (\p_1 a_2 - \p_2 a_1) \hat{e}^1 \wedge \hat{e}^2\\
		& + (\p_2 a_3 - \p_3 a_2) \hat{e}^2 \wedge \hat{e}^3\\
		& + (\p_3 a_1 - \p_1 a_3) \hat{e}^3 \wedge \hat{e}^1.
	\end{aligned}
\end{equation}

The property $\mathbf{d}^2 = 0$ is equivalent to \cref{eq:Double_wedge}.

In three dimensions, the curl, as defined in vector calculus, is obtained by using the dual operation in the same way that we obtained the cross product from the outer product,
\begin{equation}
	\vec{\nabla} \times a \Leftrightarrow - I (\nabla \wedge a) \Leftrightarrow \star \mathbf{d} \omega.
\end{equation}

\paragraph{Coderivative:} The definition of a ``divergence`" in differential forms is quite complicated and has its origin in the original description of the inner product in differential forms \footref{foot:old-inner-product}. The operation is known as \textit{coderivative} and it's expressed as
\begin{equation}
	\delta \omega = \star (\mathbf{d}\star \omega).
\end{equation}

The equivalent operation in GA is the application of the vector derivative with the inner product, which produces a generalized divergence. Being $M$ a multivector
\begin{equation}
	\nabla \cdot M = \sum_{i,j}(-1)^{j+1} \p_j M_{(i)}^{1...j...k} e_{1} \wedge ...\wedge e_{j-1}\wedge e_{j+1}\wedge... \wedge e_{k}
\end{equation}

In the case of $M$ being a vector field, $a = \sum_{i}a^i e_i$, we recover
\begin{equation}
	\nabla \cdot a = \sum_i \p_i a^i.
\end{equation}

The only difference between $\delta$ and $\nabla \cdot$ is a minus sign caused by the use of the double Hodge star in $\delta$,
\begin{equation}
	\delta \omega \Leftrightarrow - \nabla \cdot \omega.
\end{equation}

\paragraph{Laplacian:} In differential forms, the Laplacian is expressed as a combination of co- and exterior derivative. In geometric calculus, due to the associativity of the geometric product, we obtain the Laplacian operator simply by double acting with $\nabla$.
\begin{equation}
	\begin{aligned}
		(\mathbf{d} \delta + \delta \mathbf{d}) f & = \sum_{i = 1}^{3} \p_i^2 f\\
		& \Updownarrow\\
		\nabla^2 f & = (\nabla \cdot \nabla) f = \sum_{i = 1}^{3} \p_i^2 f
	\end{aligned}
\end{equation}
The second equality in the geometric calculus expression holds because $\nabla \wedge \nabla = 0$. Notice how consistent is the treatment of $\nabla$ as a vector, the outer product of a vector with itself vanishes and its self inner product produces an scalar. The Laplacian is a scalar operator in so far it doesn't change the grade of the element that it acts upon.

Moreover, $(\mathbf{d} \delta + \delta \mathbf{d})$ is only defined over scalar functions; while, $\nabla^2$ acts as a scalar operator over any multivector field.

\subsection{Identities}

This section will list a series of algebraic identities and their formulation in vector calculus, differential forms and geometric calculus.
In the cases where the cross-product appears in the vector calculus, we will omit the dual operation in geometric calculus in order to produce algebraic relationships valid in any dimensions.

Consider $f$ and $g$ scalar functions, $\sigma $ and $\omega$ differential one-forms and $a$ and $b$ vectors. $\vec{\nabla}$ will denote the vector calculus differential operator, while $\nabla$ will be the vector derivative of geometric calculus.

\begin{table}[h!]
	\centering
	\scalebox{0.8}{
		\begin{tabular}{@{}lc@{}}
			\toprule
			& Vector calculus   \\ \midrule
			(1) & $\vec{\nabla} (fg) = (\vec{\nabla}f) g + f (\vec{\nabla}g)  $   \\
			(2) & $\vec{\nabla} \cdot (f a) = (\vec{\nabla}f) \cdot a + f (\vec{\nabla} \cdot a)  $ \\
			(3) & $\vec{\nabla} \cdot (a \times b) = b \cdot (\vec{\nabla}\times a) - a \cdot (\vec{\nabla} \times b)$ \\
			(4) & $\vec{\nabla} \times(f a) = \vec{\nabla f} \times a + f (\vec{\nabla} \times a)$\\
			(5) & $a \cdot (b \times c) = c \cdot (a \times b) = b \cdot (c \times a) $ \\
			(6) & $ a \times (b \times c) = b (a \cdot c) - c (a \cdot b)$ \\
			(7) & $ (a \times b) \cdot (c \times d) = (a \cdot c)( b \cdot d) - (a \cdot d)( b \cdot c)$                                                              \\ \bottomrule
		\end{tabular}
	}\caption{List of some identities of vector calculus.}
\end{table}

\begin{table}[h!]
	\centering
	\scalebox{0.8}{
		\begin{tabular}{@{}lc@{}}
			\toprule
			& Differential forms   \\ \midrule
			(1) & $\mathbf{d}fg = (\mathbf{d}f)g + f (\mathbf{d}g) $   \\
			(2) & $\star (\mathbf{d}\star (f \omega)) = \star (\mathbf{d}f \wedge \omega) + f \star (\mathbf{d}\star  \omega)$ \\
			(3) & $\star (\mathbf{d} \star \star  (\sigma \wedge \omega)) = \star  (\omega \wedge \mathbf{d} \sigma) - \star (\sigma \wedge \mathbf{d}\omega)$ \\
			(4) & $\star (\mathbf{d}(f \omega)) = \star (\mathbf{d}f \wedge \omega) + f\star (\mathbf{d}\omega) $\\
			(5) & $\star (\sigma_1 \wedge \sigma_2 \wedge \sigma_3) = \star (\sigma_2 \wedge \sigma_3 \wedge \sigma_1) = \star (\sigma_3 \wedge \sigma_1 \wedge \sigma_2)$\\
			(6) & $\star (\sigma_1 \wedge \star (\sigma_2 \wedge \sigma_3)) = \sigma_2\star (\sigma_1 \wedge \star \sigma_3) - \sigma_3\star (\sigma_1 \wedge \star \sigma_2)$\\
			(7) & $\star (\star (\sigma_1 \wedge \sigma_2)\wedge\star (\sigma_3 \wedge \sigma_4)) = \star (\sigma_1 \wedge \sigma_3)\star (\sigma_2 \wedge \sigma_4) - \star (\sigma_1 \wedge \sigma_4)\star (\sigma_2 \wedge \sigma_3)$ \\ \bottomrule
		\end{tabular}
	}\caption{List of some identities of differential forms.}
\end{table}

\begin{table}[h!]
	\centering
	\scalebox{0.8}{
		\begin{tabular}{@{}lc@{}}
			\toprule
			& Geometric calculus  \\ \midrule
			(1) & $\nabla (fg) = (\nabla f) g + f (\nabla g)  $ \\
			(2) & $\nabla \cdot (f a) = (\nabla f) \cdot a + f \nabla \cdot a $ \\
			(3) & $\nabla \cdot (a \wedge b) = (\nabla \cdot a) b - a (\nabla \cdot b) $ \\
			(4) & $\nabla \wedge (f a) = (\nabla f) \wedge a + f (\nabla \wedge a)$\\
			(5) & $ a \wedge b \wedge c = b \wedge c \wedge a = c \wedge a \wedge b$\\
			(6) & Same as (3): $a \cdot (b \wedge c) = a \cdot b c - a \cdot c b$\\
			(7) & $(a \wedge b) \cdot (c \wedge b) = (a \cdot (b \cdot (c \wedge b))) = (a \cdot d)(b\cdot c) - (a \cdot c)(b \cdot d)$  \\ \bottomrule
		\end{tabular}
	}\caption{More identities of geometric calculus and its proofs can be found in \cite{Hitzer2013}}
\end{table}

From the form of these identities we can say that geometric calculus is a language taking the best of both worlds: from vector calculus, the geometric interpretation and the intuitiveness of the language, and from differential forms, the exterior algebra, computational power and directed integration. However, in GC, the mixture turns out to be much more than their simple addition, producing, for example, a generalization complex numbers and the powerful theorems that presented in \cref{sec:fundamental-theorem-of-calculus}.

\subsection{Theorems}

In this section we will present some common theorems in GC and how they relate to their corresponding versions in differential forms language.

\paragraph{Scalar potential:} If a vector field $E$ has null curl, $\nabla \wedge E = 0$, then $\exists$ a scalar field $\phi$ such that $-\nabla \phi = -\nabla \wedge \phi = E$.

In differential forms: If a one-form $\omega$ satisfies $\mathbf{d}\omega = 0 \Rightarrow \exists $ a 0-form $\alpha $ such that $\mathbf{d}\alpha = \omega$.

\paragraph{Vector potential:} If a bivector field $B$ has null divergence, $\nabla \cdot B = 0$. Then $\exists$ a vector field $A$ such that $\nabla \wedge A = B$.

In differential forms: If a one-form $\omega$ satisfies $\star (\mathbf{d}\star \omega) = 0$, then $\exists$ a one-form $\beta$ such that $\star \omega = \mathbf{d}\beta$ or $\omega = \star (\mathbf{d}\beta)$.

\paragraph{Decomposition theorem:} For any $k$-vector $A_k$, $\exists$ $a, b, c$ such that $A_k = \nabla \wedge a + \nabla \cdot b + c$, with $c$ satisfying $\nabla^2 c = 0$. \cite{Roberts_2022, Robson2023}

In differential forms: For any $p$-form $\omega$, $\exists$ $\alpha$, $\beta$ and $\gamma$ such that $\omega = \mathbf{d}\alpha + \delta \beta + \gamma$, being $\gamma$ harmonic: $\nabla^2 \gamma = 0$.

\paragraph{Gauss and Stokes theorems:} Gauss and Stokes's theorems are particular cases of the fundamental theorem of calculus, \cref{eq:Fundamental_Theorem}, as we showed in \cref{sec:fundamental-theorem-of-calculus}.

In differential forms, we can write these theorems using the generalized Stokes theorem and the Hodge star operator as
\begin{equation}\label{eq:Gen_Stokes}
	\int_D \mathbf{d}\omega = \int_{\p D} \omega
\end{equation}
\begin{equation}\label{eq:Gen_dual_Stokes}
	\int_D \mathbf{d}(\star \omega) = \int_{\p D} \star \omega
\end{equation}

\Cref{eq:Gen_Stokes} can be derived from \cref{eq:Fundamental_Theorem} by taking the two-dimensional case and $M$ being a scalar function $f$ and identifying on the right side the one-form $\dd x_1 f \Leftrightarrow \mathbf{d}\omega $, and on the left the exterior derivative $\dd x_2 \cdot (\nabla \wedge f) \Leftrightarrow \mathbf{d}\omega$.

\section{Electrodynamics}\label{sec:Maxwell}

We can find an elegant example of the power of geometric calculus in its description of the laws of electromagnetism\cite{Dressel2015}. Because EM is a relativistic phenomenon, this section will start presenting the Clifford space of Minkowski space-time with metric $ \eta_{\mu \nu} = (+1,-1,-1,-1)$, called $\C{1,3}$. We can summarize the relation between the differrent vector basis elements of $\C{1,3}$, $\{\g_\mu\}$, as
\begin{equation}
	\g_\mu \cdot \g_\nu = \eta_{\mu \nu}.
\end{equation}
From them, we construct the rest of the basis elements of $\C{1,3}$ from this vector basis as shown in \cref{tab:STA_Basis}.

\begin{table}[h!]
	\begin{tabular}{l | c | c | c | c | c | c }
		Scalars & 1 &  & & & \\ 
		4 vectors & $ \gamma_0 $ & $ \gamma_1 $ & $ \gamma_2 $ & $ \gamma_3 $ & & \\
		6 bivectors & $ \gamma_{10} $ & $ \gamma_{20} $ &$  \gamma_{30} $ & $ \gamma_{23} $ & $ \gamma_{31} $ & $ \gamma_{12} $\\
		4 trivectors & $ \gamma_{123} $ & $ \gamma_{230} $ & $ \gamma_{310} $ & $ \gamma_{120} $ &  \\ 
		1 Pseudoscalar &$ \gamma_{0123} $ & & & & \\
	\end{tabular}
	\caption{Basis elements of $\mathcal{C}l_{1,3}$. Notice how the scalar and the pseudo scalar, the vectors and the trivectors and the first three bivectors and the last three bivectors are dual of each other.}
	\label{tab:STA_Basis}
\end{table}

In geometric calculus, the electromagnetic field is described by the bivector field $F$, which can be decomposed as
\begin{equation}
	F =E^1 \gamma_{10}+E^2 \gamma_{20}+E^3 \gamma_{30}+\left(B^1 \gamma_{10}+B^2 \gamma_{20}+B^3 \gamma_{30}\right)I 
\end{equation}
Where we have used the convention $\g_\mu \wedge \g_\nu = \g_{\mu\nu}$ to simplify the expression and expressed the space-space bivectors as the dual of the space-time bivectors..

A feature of GA of space-time is its projection into the observer's 3-dimensional space. This projection is possible because there is an isomorphism between the even subalgebra of $\C{1,3}$ and $\C{3,0}$. We can map the space-time bivector basis elements into the vector basis of $\C{3,0}$, $\gamma_{i0} \leftrightarrow \hat{e}_i$. This map allows us to express the relativistic electromagnetic bivector $F$ in the observer's base as
\begin{equation}\label{eq:EM_3D}
	F=E^1 \hat{e}_1+E^2 \hat{e}_2+E^3 \hat{e}_3+(B^1 \hat{e}_1+B^2 \hat{e}_2+B^3 \hat{e}_3)I=\vec{E}+\vec{B} I.
\end{equation}

In the 3-dimensional space of an observer with 4-velocity $\g_0$, the electromagnetic bivector $F$ decomposes into a vector electric field, and the dual of the magnetic field. This vector-bivector description of the electric and magnetic field explains their different behavior under parity transformations and the weird classical distinction between axial and polar vectors.

Moreover, noting that $I^2 = -1$ in Minkowski space-time, we notice that \cref{eq:EM_3D} is precisely the Riemann-Silberstein vector, $\vec{E} + i \vec{B}$. The complex vector form of the electromagnetic field which has been shown to considerably simplify the treatment of certain electromagnetic systems and aid in the formulation and interpretation of QED \cite{Silberstein_1907,Bialynicki_Birula_2013}. This underlying ``complex" structure is not manifest when working with the tensor components $F^{\mu\nu}$ or the Faraday two-form $\mathbf{F}$.

Switching now to differential forms, the electromagnetic bivector is equivalent to the Faraday two-form $\mathbf{F} = \mathbf{F_{\mu\nu}} \dd x^\mu \wedge \dd x^\nu$ and Maxwell's equations reduce to the pair of equations
\begin{equation}
	\begin{aligned}
		\mathbf{d} \mathbf{F} & = 0 \leftrightarrow \mathbf{F_{[\mu\nu, \lambda]}} =0 \\
		\star \mathbf{d} \star \mathbf{F} & = \frac{4\pi}{c} \mathbf{J} \leftrightarrow \mathbf{F\indices{^{\mu\nu}_{ ,\nu}} } = \frac{4\pi}{c} \mathbb{J}^\mu.
	\end{aligned}
\end{equation}
Where $[\mu\nu, \lambda]$ denotes skew-symmetry in all indices and $\mathbf{J}$ is a one-form representing the currents and charges.

In geometric calculus, Maxwell's equations are reunited in a single equation, the simplest differential equation that can form with a bivector field. Producing the most compact form of Maxwell's equations in literature.
\begin{equation}\label{eq:GA_Maxwell}
	\nabla F = \frac{4\pi}{c}j.
\end{equation}
Where $j = \rho \g_0 + \jmath^i \g_i$ is the 4-vector corresponding to the usual 4-current density.

To recover vector Maxwell's equations from \cref{eq:GA_Maxwell}, we simply need to perform the aforementioned time-split for $F$, \cref{eq:EM_3D}, for $\nabla =  \g_0(\p_0 + \vec{\nabla})$ and expand.
\begin{equation}\label{eq:Maxwell_3D}
	\begin{aligned}
		\nabla F(x) & = \nabla \cdot F(x) + \nabla \wedge F(x)\\
		& = \gamma_0(\partial_0 +\vec{\nabla})(\vec{E} + \vec{B}I)\\
		& = \gamma_0 \left[ \vec{\nabla}\cdot \vec{E} + \partial_0 \vec{E} -\vec{\nabla}\times \vec{B}\right]\\
		& + \gamma_0 \left[ \vec{\nabla} \cdot \vec{B} + \partial_0 \vec{B} + \vec{\nabla} \times \vec{E}\right]I.
	\end{aligned}
\end{equation}

We recover the classical form of Maxwell's equations by equating the vector part of \cref{eq:Maxwell_3D} to the vector $j$ and its bivector part to 0. The presence of $\g_0$ on the left emphasizes that the expression is only valid for the observer with 4-velocity $\g_0$.

The expression of Maxwell's equation in geometric calculus is more than an aesthetic exercise. The invertibility of the $\nabla$ function, which is Green's function, permits to invert \cref{eq:GA_Maxwell} and to obtain the fields directly from the sources without passing through second-order derivatives. Meaning that Helmholtz's equation is first order, and Huygens' principle of re-radiation is directly applicable and evident from the solution; see \cite{Gull_Lasenby_Doran_1993, Doran2013} for details.

Another remarkable feature of GC is that we can obtain Maxwell's equations as a particular case of Dirac's equation \cite{Gull_Lasenby_Doran_1993}. This is an excellent example of the benefits that carries having a unified formalism for physics, where we can apply the same mathematical techniques to study different phenomena. 

One finds similar elegance and compactness when describing other elements of electromagnetism, such as potentials, the Poynting flux or the Lorentz force. Where the geometric product permits the combination of linear and angular elements into single equations. E.g. work and torque, or linear and angular momentum.

A comprehensive and precise exposition of electromagnetism with geometric algebra can be found in \cite{Dressel2015}. Its first pages include a remarkable list of 33 characteristic advantages of geometric calculus to describe electromagnetism.

\section{General Relativity and Cartan's formalism}

Another realm where differential forms are widely used is in General Relativity (GR). Compared with tensor formalism, differential forms together with  Cartan's structure equation provide an efficient way of obtaining the connection elements.

In a past article\cite{Perez2024}, we presented a new formalism for working in GR which we called tetrad-GA. On it, we exploit the orthonormality properties of tetrads together with the geometric interpretation of GA to describe GR in a concise and intuitive manner.

The treatment of GR with differential forms is very similar to the one with the tetrad-GA formalism and many of the points that we presented until now also apply here. Therefore, in this section we will cover only further aspects that are characteristic of GR.

\subsection{The frames}

In both formalisms the use orthonormal frames is wide spread, which are denoted as $\{\theta_{\hat{a}}\}$ in the differential forms language, and $\{\g_m\}$ in the tetrad-GA formalism. They relate to the coordinate frame, $\{\p_\mu\}$ in differential forms, and $\{g_\mu\}$ in tetrad-GA, by a transformation called vierbein $e\indices{^m_\mu}$
\begin{equation}
		g_\mu = e\indices{^m_\mu} \g_m \quad \leftrightarrow \quad \p_\mu = e\indices{^{\hat{a}}_\mu}\theta_{\hat{a}}.
\end{equation}

The vierbein has an inverse, denoted by $e\indices{_m^\mu} $ and is defined by
\begin{equation}\label{eq:Vierbein_Inv_Def}
	\T{e}{^m_\nu}\T{e}{_m^\mu} = \delta^\mu_\nu, \quad 	\T{e}{^m_\mu}\T{e}{_n^\mu} = \delta^m_n.
\end{equation}
The inverse vierbein relates the reciprocal frames
\begin{equation}\label{eq:non-important-eq}
	g^\mu = \T{e}{_m^\mu}\g^m \quad \leftrightarrow \quad \dd x^\mu = e\indices{_{\hat{a}}^\mu}\theta^{\hat{a}}
\end{equation}
and allows to perform the inverse transformations, from the tetrad back to the coordinate frame. The coordinate one-forms are denoted as $\{\dd x^\mu\}$ in differential forms while in GA we denoted with $g^\mu$. In \cref{eq:non-important-eq} we used the musical isomorphism to define $\theta_{\hat{a}} = \eta_{\hat{a}\hat{m}} \left(\theta^{\hat{m}}\right)^\sharp$.

Even though both formalism use orthonormal frames, only the tetrad-GA formalism fully exploits its power by making an explicit separation of the degrees of freedom related to coordinates from those related to the local frame, and by allowing direct treatment of spinors in curved space-times\cite{gu_theory_2021}.

The choice of frame in the tetrad-GA formalism can be considered as a special case of the tetrads used in the ADM formalism. While in the tetrad-GA we normally use the frame of an inertial observer, whose local metric is Minkowskian, 
\begin{equation}
	\g_{mn} = \pm \begin{pmatrix}
		1 & 0 & 0 & 0\\
		0 & -1 & 0 & 0\\
		0 & 0 & -1 & 0\\
		0 & 0 & 0 & -1\\
	\end{pmatrix}
\end{equation}
in the ADM formalism the only requirement is the time axis to be orthonormal to the spatial axis, so we can write the local metric of the ADM frame as
\begin{equation}
	\g_{mn} = \pm \begin{pmatrix}
	1 & 0\\
	0 & \gamma_{ij}
	\end{pmatrix}.
\end{equation}
Being $\{\gamma_{ij}\}$ the metric of the spatial axes.

The traditional approach to the ADM works with 
\begin{equation}
	\g_{ij} = \begin{pmatrix}
		-1 & 0 & 0\\
		0 & -1 & 0\\
		0 & 0 & -1\\
	\end{pmatrix}
\end{equation}
which falls back to the same tetrad choice of inertal observers that we use in the tetrad-GA formalism. However, nothing impedes us from making other choice for $\g_{ij}$ in the ADM formalism, nor from chosing a different $\g_{mn}$ the tetrad-GA to other tetrad frame.

\subsection{Connection elements}

Given an arbitrary basis for vector fields $\{e_a\}$ and for one-forms $\{\theta^a\}$, the connection coefficients one-forms $\omega\indices{^a_b}$ are defined as 
\begin{equation}\label{eq:def-connection}
	\nabla_u e_a = \omega\indices{^b_a}(u) e_b,
\end{equation}

In a similar manner, in the tetrad-GA formalism we write \cref{eq:def-connection} as
\begin{equation}\label{eq:connection-ga}
	u \cdot D e_a = D_u e_a = \frac{1}{2}[\omega(u), e_a] = \frac{1}{2}\omega(u) \cdot e_a.
\end{equation}
Being $D$ the covariant vector operator, and in the last step we used the fact that $e_a$ is a vector. The $1/2$ factor stems from the fact that the infinitesimal transformation of an orthonormal frame when parallel transported needs to be a Lorentz transformation, which can be expressed as the exponential of a bivector with a $1/2$ factor that drops down when it is series expanded.

In spite of their similarities, $\omega\indices{^b_a}(u)$ and $\omega(u)$ are different objects. $\omega\indices{^b_a}(u)$ is a scalar without any prior symmetries, while $\omega(u)$ is a bivector. That means that $\omega(u)$ is anti-symmetric and we can decompose it in the orthonormal frame $\{\g^m\}$,
\begin{equation}
	\omega(\g_l) = \omega_l = \omega_{mnl} \gamma^m \wedge \g^n
\end{equation}
The components $\omega_{mnl} = -\omega_{nml}$ are called the \textit{Ricci rotation coefficients}. 

The Ricci rotation coefficients can also be obtained from the connection one-forms when we express them in an orthonormal frame of the cotangent space $\theta^{\hat{a}}$
\begin{equation}
	\omega\indices{^{\hat{a}}_{\hat{b}}} = \omega\indices{^{\hat{a}}_{\hat{b}\hat{c}}}\theta^{\hat{c}}.
\end{equation}
In this case
\begin{equation}
	\omega\indices{_{\hat{a}\hat{b}\hat{c}}} = \omega_{mnl}.
\end{equation}

The bivectorial nature of the connection coefficients in the tetrad-GA formalism gives them a neat geometrical interpretation. When expressed in the orthonormal frame of an inertial observer, $\omega(g_\mu)$ is the generator of the Lorentz transformation that the observer experiences when parallel transported in the $\mu$-direction.

When working in torsion-free spaces the connection one-forms are usually obtained via \textit{Cartan's first structure equation}
\begin{equation}\label{eq:Cartan-1}
	\mathbf{d}\theta^{b} = -\omega\indices{^b_a} \wedge \theta^a,
\end{equation}
with the \textit{guess and check method}.

In the tetrad-GA case, we obtain the the connection bivectors using
\begin{equation}\label{eq:Connection-coefficients}
	\omega_\mu = \frac{1}{2}\left(g^\lambda \wedge \nabla g_{\mu\lambda} + g_\alpha \wedge \p_\mu g^\alpha\right).
\end{equation}
Where the last term is computed as  $\p_\mu g^\alpha = \g^m\p_\mu \T{e}{_m^\alpha}$.

In the case of having a diagonal metric, \cref{eq:Connection-coefficients} reduces to
\begin{equation}\label{eq:Connection-coefficients-diagonal}
	\omega_\mu = \frac{1}{2}g^\mu \wedge \nabla g_{\mu\mu} = \frac{1}{2}g^\mu \wedge g^\nu \p_\nu g_{\mu\mu},
\end{equation}
without summation over the $\mu$ indices. \Cref{eq:Connection-coefficients-diagonal} is extremely simple to calculate. With a maximum of 16 derivatives, it is the fastest and most straight-forward way of obtaining the connection elements known to the authors.

In summary, the connection coefficients in differential forms and the tetrad-GA are very similar objects with minor variations often due to the chosen basis. The main differences are two: their geometric interpretation and the method of calculation.

\subsection{Curvature two-form}

In differential forms, curvature of the manifold is characterized by the curvature two-form $\Omega\indices{^a_b}$, given by
\begin{equation}
	\Omega\indices{^a_b} = \mathbf{d}\omega\indices{^a_b} + \omega\indices{^a_c}\wedge\omega\indices{^c_b}.
\end{equation}
We can relate it to the components of the Riemann tensor in the tetrad frame as
\begin{equation}
	\Omega\indices{^a_b} = \frac{1}{2}R\indices{^a_{bcd}}\omega^c \wedge \omega^d,
\end{equation}
and it is completely equivalent to the Riemann tensor bivector in the tetrad-GA formalism
\begin{equation}
	R_{ab} = e\indices{_a^\alpha} e\indices{_b^\beta} R_{\alpha\beta} = e\indices{_a^\alpha} e\indices{_b^\beta} \hf R_{\mu\nu c d} \gamma^c \wedge \gamma^d
\end{equation}
with its first index raised
\begin{equation}
	\Omega\indices{^a_b} = \eta^{a c}R_{cb}.
\end{equation}

The only difference between both objects come their interpretation. In the tetrad-GA formalism the Riemann tensor is a linear function from bivectors to bivectors,
\begin{equation}
	\begin{aligned}
		\mathbf{R}: \Lambda^2(\mathcal{V}) &\rightarrow \Lambda^2(\mathcal{V})\\
		B = B^{\mu\nu}g_\mu \wedge g_\nu \in \Lambda^2(\mathcal{V}) & \mapsto \mathbf{R}(B) = B^{\mu\nu} \mathbf{R}_{\mu\nu} \in \Lambda^2(\mathcal{V}).
	\end{aligned}
\end{equation}

\begin{figure}
	\centering
	\def\svgwidth{0.7\textwidth}
	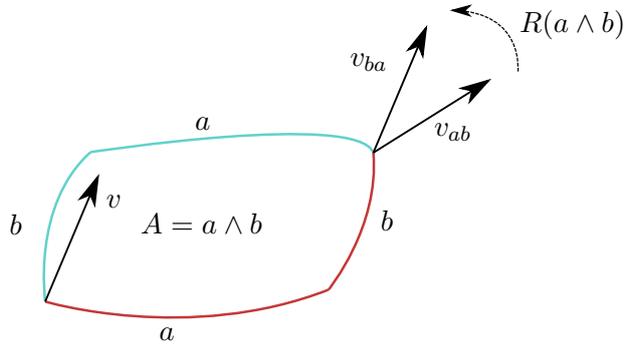
	\caption{Representation of the effect of transporting vector $ v $ to the same point through two different paths. When transported through the red path, $ a \rightarrow b $, the resulting vector is $ v_{ab} $. When transported through the blue path, $ b \rightarrow a $, the resulting vector is $ v_{ba} $. $ v_{ab} $ and $ v_{ba} $ are related by a rotation that is a function of the area spanned between the paths, $ A = a \wedge b $. That function is called the Riemann tensor.} 
	\label{fig:Curvature}
\end{figure}

Its role is connecting a bivector area, $g_\mu \wedge g_\nu$, with a bivector which is the generator of the rotation that a vector experiences when transported around its perimeter, see \cref{fig:Curvature}. Hence, it's more intuitive to use the mixed indices form, $R(g_\mu\wedge g_\nu) = 1/2 R_{\mu\nu c d} \gamma^c \wedge \gamma^d$: coordinate indices for the coordinate area and tetrad indices for the local rotation.

In this sense, we can say that the tetrad-GA formalism has a clearer geometric interpretation than the curvature two-form, which is usually said to be related to curvature, but rarely explicitly explained how.

\section{Conclusion}

In this article, we have presented the basic elements of Geometric Algebra (GA) and Geometric Calculus (GC) and compared them with the language of differential forms. The identification of differential forms with elements of GC depends on their usage and can correspond to basis elements of the reciprocal space or differential  k -vectors.

In many aspects, differential forms and GC are similar and offer considerable advantages over vector or tensor calculus. However, GC presents some distinct advantages over differential forms:

\begin{enumerate}
	\item Quantities and operations have easier interpretation in GC.
	\item The unification of the inner and outer product in the geometric product provides compact and unified expressions and integral theorems.
	\item Integral theorems in GC are more general and contain more information than the generalized Stokes’ theorem of differential forms.
	\item There exists an isomorphism between even subalgebras and complex numbers in any dimension, which extends analytic function theory to spaces of arbitrary dimensions.
	\item The unification of Cauchy’s theorem of complex analysis with other integral theorems and its geometric interpretation is remarkable and unique to GC.
	\item  GC offers considerable advantages over differential forms when dealing with rotations or spinors.
\end{enumerate}

Conversely, differential forms benefit from an extensive body of literature. The formalism is more mature, with numerous applications and textbooks available. GA, in contrast, suffers from a lack of didactic material, which can result in a steep learning curve.

As a specific example, we presented the fundamental equations of electrodynamics in both formalisms. In differential forms, Maxwell’s equations reduce to a pair of differential equations over the Faraday two-form. GC, however, achieves a greater unification of Maxwell’s equations into a single differential equation that can be solved directly, without second-order derivatives. The GC electromagnetic bivector, when projected into the observer’s frame, can be written in the form of the Riemann-Silberstein vector, providing a neat geometric explanation and interpretation.

Furthermore, we found that relativistic electromagnetic fields are better understood through the GC interpretation of lines and planes compared to the surfaces and tubes pierced by vectors offered by differential forms.

Finally, we connected Cartan’s formalism of General Relativity (GR) with our tetrad-GA formalism, reaching similar conclusions as in the case of electrodynamics. Both treatments are capable of dealing with GR, but the tetrad-GA formalism better reflects the symmetries of the objects and provides an easier geometric understanding of processes.

There is no doubt about the power of differential forms, and their inclusion of exterior calculus marked a significant improvement over the vector and tensor treatment. However, in comparison with GC, differential forms lack several characteristics with deeper consequences, such as a unified product and mixed grade elements. Additionally, the traditional notation of differential forms can be somewhat murky and difficult to grasp.

There is a modern trend in physics to emphasize its geometric content more, as noted in \cite{thorne_modern_2017}. Based on the comparison presented in this article, we encourage students and researchers to learn GC and apply it to various geometric problems. We are confident they will be impressed by its power and efficiency.

\section*{Acknowledgments}

Pablo Bañón Pérez and Maarten DeKieviet would like to thank the Vector-Stiftung, in the framework of the MINT innovation program, and the Heideberg Graduate School For Physics for their financial support

\bibliographystyle{unsrt}
\bibliography{Biblio}

\end{document}